# Method "Ethanol as Internal Standard" for determination of volatile compounds in alcohol products by gas chromatography in daily practice

[a]Siarhei V. Charapitsa, [a]Svetlana N. Sytova,
[a]Andrey A. Makhomet, [b]Tatiana I. Guguchkina,
[b]Mikhail G. Markovsky, [b]Yurii F. Yakuba, [c]Yurii N. Kotov

[a]Research Institute for Nuclear Problems of Belarusian State University
POB 220030, Bobryiskaya Str., 11, Minsk, Belarus

[b]Federal State Budgetary Scientific Institution North Caucasian Regional Research Institute
of Horticulture and Viticulture, 40 Let Pobedy Str., 39, Krasnodar 350901, Russia

[c]Branch of Joint Stock Company "Rosspirtprom" Wine and Distillery Plant "Cheboksary",
POB 428018, K. Ivanov Str., 63, Cheboksary, Russia

Corresponding author: Dr. Siarhei V. Charapitsa; e-mail: chere@inp.bsu.by;
svcharapitsa@tut.by ; Research Institute for Nuclear Problems of Belarusian State University,
POB 220030, Bobryiskaya Str., 11, Minsk, Belarus

Recently proposed new method "Ethanol as Internal Standard" for determination of volatile compounds in alcohol products by gas chromatography (GC) is investigated from different sides including method testing on prepared standard solutions like cognac and brandy, different ethanol-water solutions and certified reference material CRM LGC5100 Whisky-Congeners. Analysis of obtained results of experimental study from four different laboratories shows that relative bias between the experimentally measured concentrations calculated in accordance with proposed method and the values of concentrations assigned during the preparation by gravimetric method for all analyzed compounds does not exceed 10 %. It is shown that relative response factors (RRF) between analyzed volatile compounds and ethanol do not depend on time of analysis and are constant for every model of GC. It is shown the possibility to use predetermined RRF in daily practice of testing laboratories and to implement this new method in the international standards of measurement procedure.

## Introduction

Daily consumption of alcoholic beverages (cognac, brandy, whisky, vodka, liqueur, wine, cider, beer, etc.) worldwide creates the need for rapid and qualitative determination of volatile compounds in such products in analytical and commercial laboratories around the world. The quantitation of ethanol in these products entails the payment of taxes to the state budget. So, numerous analytical methods for determination of alcohol product composition are developed starting with the middle of the twentieth century. A long list (Wang et al., 2003; Wang et al., 2004) of different methods of determination of volatile compounds argues to this. The most popular and widely used method is GC with adding the Internal Standard (IS) in analysed sample. As IS one uses acetonitrile, 2-pentanol, 4-methyl-1-pentanol or other compounds (Wang et al., 2003; Brill & Wagner, 2012; MacNamara et al., 2010; Kostik et al., 2013). Such compounds do not appear in initial solution by fermentation but only by special adding in the laboratory. Other approach in GC is External Standard (ES) method, well-known for some decades (Patent US, 1975). In ES method, known data of some calibration standards and sample data are combined to quantitative report. There are some combinations of both methods (MacNamara et al., 2010).

We proposed (Charapitsa et al., 2013) new method of direct determination of volatile compounds in alcohol products. Its quintessence is to use ethanol as IS for the analysis of alcohol products in order to increase the accuracy of measurements and to obviate the need for the IS addition in the analysing sample. Indeed, ethanol is contained already in the sample. Full theoretical background of the method and different sides of its experimental examination are given (Charapitsa et al., 2012; Charapitsa et al., 2013; Charapitsa et al., 2014; Charapitsa et al., 2016). The first ideas of using main component (solvent) as internal standard for chromatographic quantitative analysis of impurities were formulated in 2003 (Charapitsa et al., 2003). On the basis of proposed in cited above papers on-line calculator AlcoDrinks on correct determination of volatile compounds, including ethanol, in alcohol products was developed. It can be found at address http://inp.bsu.by/calculator/vcalc.html and it work can be tested by interested users.

This article is devoted to the further comprehensive verification carried out in four different analytical laboratories. These experiments were performed with standard working ethanol-water solutions and with the certified reference material CRM LGC5100 Whisky-Congeners. There was done comparison of the results obtained by the new method with "traditional" methods: IS method using cyclohexanol as IS and ES method.

Thus, the new method "Ethanol as Internal Standard" shows itself really fast and cheap. Nowadays, the testing laboratories are equipped with modern GC for the analysis of alcohol-containing products. These current-technology gas chromatographs have a linear range of registration of seven orders of magnitude. Analysis of alcohol products in the new method consists in the procedure of determining the relative ratio of the detector response of analysed compounds with respect to detector response of ethanol by standard working solutions and then the subsequent use of these RRF in the calculation of concentration of analysed volatile compounds. It should be noted that for modern chromatographs coefficients RRF are stable and can be tabulated (ASTM, 2009; Cicchetti et al., 2008). The last fact is confirmed by numerous papers (Kolai & Balla, 2002; Rome & McIntyre, 2012), where it is shown that RRF are similar for different instruments because of its dependency only on the chemical reaction in the GC flame. So, we propose to use given in this article our average RRF for analysis on any GC within the frame of new method "Ethanol as IS" for evaluation measurements. For a given specific GC these RRF can be corrected in accordance with proposed theoretical background and experimental results of standard working solutions obtained on this GC.

**Theoretical**

Full theoretical background of the proposed method is given (Charapitsa, 2013). Let us point out here only one item concerning RRF. The mass concentration of ethanol $C_{ethanol}$ in absolute (anhydrous) alcohol (AA) is well known and it is equal to $\rho_{ethanol} = 789300$ mg/L. For the quantitative calculation of the volatile compounds in the test sample one can use the internal standard (IS) method. Let us consider an ethanol-containing sample with $i=1,2,...N$ volatile compounds. The value of the concentration of the $i$-th compound in the test sample in accordance with IS method can be described by the following formulas:

$$C_i = C_{IS} \cdot RRF_i \; \frac{A_i}{A_{IS}} \quad [\text{mg/L (AA)}], \tag{1}$$

$$RRF_i = RF_i / RF_{IS} = \frac{C_i^{st}}{A_i^{st}} / \frac{C_{IS}^{st}}{A_{IS}^{st}} , \tag{2}$$

where $C_{IS}$ is the concentration of IS in analyzing sample, expressed in mg per litre of absolute alcohol; $A_i$ and $A_{IS}$ are the peak areas of the $i$-th compound and IS, respectively; $RF_i$ and $RF_{IS}$ are the detector responses of the $i$-th compound and of IS, respectively; $C_i^{st}$ and $C_{IS}^{st}$ are the concentrations of the $i$-th compound and IS in standard working solution, expressed in mg per litre of absolute alcohol, respectively; $A_i^{st}$ and $A_{IS}^{st}$ are the peak areas of the *i*-th compound and IS in standard working solution, respectively; $RRF_i$ is relative response factor for the *i*-th compound.

Since ethanol is present in the sample and its concentration in the anhydrous alcohol is well known, the expressions (1) and (2) can be written as follows:

$$C_i = \rho_{ethanol} \cdot RRF_i \, \frac{A_i}{A_{ethanol}} \quad [\text{mg/L (AA)}], \qquad (3)$$

$$RRF_i = \frac{A_{ethanol}^{st} \cdot C_i^{st}}{A_i^{st} \cdot \rho_{ethanol}}, \qquad (4)$$

where $A_{ethanol}$ and $A_{ethanol}^{st}$ are the peak areas of ethanol in test sample and in the standard working solution, respectively.

The absolute response factors $RF_i$ and $RF_{IS}$ are changed from day to day and from one GC to another. At the same time relative response factors $RRF_i$ (4) are constant and not affected by changes in time of analysis or GC instruments from one model.

## Experimental

### *Laboratories and equipments*

To continue comprehensive examination of the proposed method there were planned and carried out experimental studies with different GC in the following laboratories:

1. Laboratory of Analytical Research of Research Institute for Nuclear Problems of Belarusian State University (INP BSU) (Minsk, Belarus), GC Chromatec-Crystal5000 (JSC "Chromatec", Russia);
2. Centre of Joint Use of Analytical Equipment (NCRRIH&V-1) of Federal State Budgetary Scientific Institution North Caucasian Regional Research Institute of Horticulture and Viticulture (Krasnodar, Russia), GC Chromatec-Crystal2000M (JSC "Chromatec", Russia);
3. Scientific Centre "Vinodelie" (NCRRIH&V-2) of Federal State Budgetary Scientific Institution North Caucasian Regional Research Institute of Horticulture and Viticulture (Krasnodar, Russia) (Krasnodar, Russia), GC Chromatec-Crystal2000M (JSC "Chromatec", Russia);
4. Control Laboratory of Branch of Joint Stock Company "Rosspirtprom" Wine and Distillery Plant "Cheboksary" (CLCheb) (Cheboksary, Russia), GC HP6890 (Agilent Technologies, USA).

All mentioned GC were equipped with flame ionization detector (FID). Parameters of operating conditions of all GC in the experiments are given in Table 1.

*Chemicals*

All individual standard compounds were purchased from Sigma-Fluka-Aldrich (Germany). The standard working solutions were prepared by adding the individual standard high-grade compounds to the ethanol-water mixture (96:4) by gravimetric method according to ASTM D 4307 recommendations (ASTM, 2007). Calculated parameters of the prepared standard working solutions are given in the next section. Certified reference material CRM LGC5100 Whisky-Congeners was purchased from LGC Standards Sp. z o. o. (Poland).

**Table 1.** Parameters of GC operating conditions

| Laboratory | INP BSU | NCRRIH&V-1 | NCRRIH&V-2 | CLCheb |
|---|---|---|---|---|
| GC | Chromatec-Crystal 5000 | Chromatec-Crystal 2000M | Chromatec-Crystal 2000M | HP6890 |
| split/splitless injector | Yes | Yes | Yes | Yes |
| Software | Unichrom (New Analytical Systems Ltd., Belarus) | Chromatec-Analytic (JSC "Chromatec", Russia) | Chromatec-Analytic (JSC "Chromatec", Russia) | Unichrom (New Analytical Systems Ltd., Belarus) |
| capillary column | Rt-Wax, 60 m × 0.53 mm, phase thickness 1 μm (Restek, Bellefonte, PA) | HP-FFAP, 50 m × 0.53 mm, phase thickness 0.5 μm (Agilent Technologies, USA) | HP-FFAP, 50 m × 0.53 mm, phase thickness 0.5 μm (Agilent Technologies, USA) | ZB-FFAP, 50 m × 0.32 mm, phase thickness 0.5 μm (Agilent Technologies, USA) |
| oven temperature | the initial isotherm at 75°C (9 min), raised to 155°C at a rate of 7°C/min, with final isotherm of 155°C (2.6 min) | the initial isotherm at 70°C (7 min), raised to 170°C at a rate of 10°C/min, with final isotherm of 170°C (3 min). | the initial isotherm at 75°C (7 min), raised to 170°C at a rate of 7°C/min with final isotherm of 170°C (3 min). | the initial isotherm at 75 °C (3 min), raised to 170°C at a rate of 10°C/min with final isotherm of 170°C (3 min). |
| carrier gas | nitrogen | nitrogen | nitrogen | hydrogen |
| Gas flow | 2.44 mL/min | 1.25 mL/min | 1.0 mL/min | 2.7 mL/min |
| injector temperature | 160°C | 150°C | 180°C | 170°C |
| detector temperature | 200°C | 180°C | 200°C | 220°C |
| injector volume | 0.5 μL | 1.0 μL | 0.5 μL | 1.0 μL |
| split ratio | 1:20 | 1:30 | 1:20 | 1:20 |

## Results and discussion

### *Experiments with standard working solutions like cognac and brandy*

The first experiments have been performed in NCRRIH&V-2 on the GC Chromatec-Crystal2000M. Content of volatile compounds in the first series of the prepared standard solutions was chosen like in cognac and brandy products. Standard working solutions A(5000), B(1000), C(100), D(10) and E(2) were prepared by gravimetric method. The initial standard solution A(5000) was prepared by adding the individual compounds to high-grade ethanol. A 100 mL volumetric flask and OHAUS PA214C analytical balance with a margin error of measurement not worse than 0.2 mg were used for preparation of the initial standard solution A(5000) with a mass concentration in the range 5000 mg/L of absolute alcohol. Fifty milliliters of ethanol was added into the flask and weighed. Then 0.5 mL of individual compounds were added into the flask. The exact weight of each added compound was recorded. Ethanol was added up to the label and weighed. In calculations, it was considered that the following impurities were present in the initial ethanol (rectified ethyl alcohol): acetaldehyde 0.162 mg per 1 L of AA; methanol 2.53 mg per 1 L of AA and 2-propanol 1.35 mg per 1 L of AA. Subsequent standard solutions from B(1000) till D(10) were prepared by adding solution A(5000) to high-grade ethanol in the following ratios: for B(1000): 1 part A(5000) to 4 parts ethanol; for C(100) the ratio was 1:49; for D(10) 1:500. The standard solution E(2) was obtained by dilution of C(100) with ethanol in proportion 1:50.

**Table 2.** Sample A(5000). Comparison of experimentally measured concentrations of analyzed volatile compounds in standard solutions obtained in NCRRIH&V-2 by three methods: cyclohexanol as IS, the ES method and method of using ethanol as IS with initial concentration according to gravimetric method

| | acetaldehyde | methyl acetate | ethyl acetate | methanol | 2-propanol | ethanol | 1-propanol | isobutyl alcohol | isoamyl acetate | 1-butanol | isoamyl alcohol | ethyl hexanoate | cyclohexanol | ethyl octanoate | ethyl decanoate | benzyl alcohol | phenylethanol |
|---|---|---|---|---|---|---|---|---|---|---|---|---|---|---|---|---|---|
| | | | | | | | | | **Gravimetric method** | | | | | | | | |
| Concentration, mg /L (AA) | 5870 | 5642 | 5681 | 5894 | 5619 | 789300 | 5523 | 5556 | 5545 | 5597 | 5657 | 5580 | 111 | 5454 | 5745 | 5642 | 5766 |
| amount, (0,5 mcl Split=20) pg | 128,7 | 123,7 | 124,6 | 129,2 | 123,2 | 17307 | 121,1 | 121,8 | 121,6 | 122,7 | 124,1 | 122,4 | 2,4 | 119,6 | 126,0 | 123,7 | 126,4 |
| response x10, pC | 264 | 211 | 304 | 254 | 382 | 45821 | 451 | 539 | 461 | 513 | 548 | 484 | 12 | 511 | 544 | 465 | 608 |
| | | | | | | | | | **Cyclohesanol as IS** | | | | | | | | |
| Concentration, mg /L (AA) | 5870 | 5597 | 5679 | 5883 | 5630 | 790426 | 5529 | 5567 | 5612 | 5565 | 5677 | 5603 | 111 | 5470 | 5751 | 5660 | 5780 |
| repeat, % | 0,5 | 2,1 | 1,9 | 1,5 | 0,9 | 1,9 | 0,6 | 0,9 | 0,6 | 0,5 | 0,4 | 0,3 | 0,0 | 0,2 | 0,6 | 0,3 | 0,4 |
| relative bias, % | 0,0 | -0,8 | 0,0 | -0,2 | 0,2 | 0,1 | 0,1 | 0,2 | 0,3 | 0,4 | 0,3 | 0,4 | 0,0 | 0,3 | 0,1 | 0,3 | 0,2 |
| | | | | | | | | | **ES** | | | | | | | | |
| Concentration, mg /L (AA) | 5694 | 5415 | 5499 | 5692 | 5474 | 657525 | 5362 | 5398 | 5445 | 5397 | 5510 | 5435 | 91 | 5304 | 5577 | 5502 | 5617 |
| repeat, % | 11,7 | 13,3 | 13,1 | 12,7 | 12,1 | 13,1 | 11,8 | 12,1 | 11,7 | 11,7 | 11,6 | 11,5 | 11,2 | 11,4 | 11,8 | 10,9 | 10,8 |
| relative bias, % | -3,0 | -4,0 | -3,2 | -3,4 | -2,6 | -16,7 | -2,9 | -2,8 | -2,7 | -2,7 | -2,6 | -2,6 | -17,3 | -2,7 | -2,9 | -2,5 | -2,6 |
| | | | | | | | | | **Ethanol as IS** | | | | | | | | |
| Concentration, mg /L (AA) | 5878 | 5605 | 5688 | 5891 | 5638 | 789300 | 5537 | 5574 | 5620 | 5573 | 5685 | 5610 | 114 | 5477 | 5758 | 5668 | 5788 |
| repeat, % | 1,4 | 0,3 | 0,0 | 0,4 | 1,0 | 0,0 | 1,2 | 1,0 | 1,3 | 1,4 | 1,5 | 1,6 | 1,9 | 1,6 | 1,3 | 2,2 | 2,3 |
| relative bias, % | 0,1 | -0,6 | 0,1 | -0,1 | 0,3 | 0,0 | 0,2 | 0,3 | 0,4 | 0,5 | 0,5 | 0,5 | 3,5 | 0,4 | 0,2 | 0,5 | 0,4 |

**Table 3.** Sample B(1000). Comparison of experimentally measured concentrations of analyzed volatile compounds in standard solutions obtained in NCRRIH&V-2 by three methods: cyclohexanol as IS, the ES method and method of using ethanol as IS with initial concentration according to the gravimetric method

| | acetaldehyde | methyl acetate | ethyl acetate | methanol | 2-propanol | ethanol | 1-propanol | isobutyl alcohol | isoamyl acetate | 1-butanol | isoamyl alcohol | ethyl hexanoate | cyclohexanol | ethyl octanoate | ethyl decanoate | benzyl alcohol | phenylethanol |
|---|---|---|---|---|---|---|---|---|---|---|---|---|---|---|---|---|---|
| **Gravimetric method** | | | | | | | | | | | | | | | | | |
| Concentration, mg /L (AA) | 1094 | 1051 | 1058 | 1119 | 1047 | 789300 | 1029 | 1035 | 1033 | 1043 | 1054 | 1040 | 114 | 1016 | 1070 | 1051 | 1074 |
| amount, (0,5 mcl Split=20) pg | 25,8 | 24,8 | 25,0 | 26,4 | 24,7 | 18613 | 24,3 | 24,4 | 24,4 | 24,6 | 24,9 | 24,5 | 2,7 | 24,0 | 25,2 | 24,8 | 25,3 |
| response x10, pC | 56,8 | 45,3 | 65,5 | 56,5 | 81,2 | 55335 | 97,7 | 117 | 100 | 111 | 118 | 105 | 15,1 | 112 | 119 | 98,8 | 129 |
| **Cyclohesanol as IS** | | | | | | | | | | | | | | | | | |
| Concentration, mg /L (AA) | 1042 | 994,3 | 1006,9 | 1075,8 | 988,0 | 784869 | 990,2 | 997,9 | 1003 | 992,0 | 1013 | 1005 | 117,4 | 993,6 | 1042 | 1001 | 1021 |
| repeat, % | 0,4 | 1,4 | 0,3 | 0,2 | 0,0 | 0,1 | 0,1 | 0,1 | 0,1 | 0,5 | 0,1 | 0,3 | 0,0 | 0,0 | 0,5 | 0,4 | 0,2 |
| relative bias, % | -4,7 | -5,4 | -4,9 | -3,9 | -5,6 | -0,6 | -3,8 | -3,6 | -3,8 | -4,0 | -3,9 | -3,3 | 0,0 | -2,2 | -2,7 | -4,8 | -5,0 |
| **ES** | | | | | | | | | | | | | | | | | |
| Concentration, mg /L (AA) | 1157 | 1101 | 1115 | 1191 | 1099 | 746942 | 1099 | 1107 | 1114 | 1101 | 1125 | 1116 | 111 | 1103 | 1156 | 1113 | 1136 |
| repeat, % | 0,7 | 2,6 | 1,5 | 1,4 | 1,2 | 1,3 | 1,1 | 1,1 | 1,0 | 0,7 | 1,0 | 0,8 | 1,2 | 1,2 | 1,7 | 1,6 | 1,0 |
| relative bias, % | 5,8 | 4,7 | 5,4 | 6,4 | 5,0 | -5,4 | 6,8 | 7,0 | 6,8 | 6,6 | 6,8 | 7,3 | -5,3 | 8,5 | 8,0 | 5,9 | 5,7 |
| **Ethanol as IS** | | | | | | | | | | | | | | | | | |
| Concentration, mg /L (AA) | 1051 | 1003 | 1016 | 1085 | 996 | 789300 | 999 | 1006 | 1012 | 1000 | 1022 | 1014 | 122 | 1002 | 1051 | 1009 | 1029 |
| repeat, % | 0,6 | 1,3 | 0,2 | 0,1 | 0,1 | 0,0 | 0,2 | 0,2 | 0,3 | 0,6 | 0,3 | 0,4 | 0,1 | 0,1 | 0,4 | 0,3 | 0,3 |
| relative bias, % | -3,9 | -4,6 | -4,0 | -3,0 | -4,8 | 0,0 | -3,0 | -2,8 | -3,0 | -3,2 | -3,1 | -2,5 | 4,2 | -1,4 | -1,8 | -4,0 | -4,2 |

**Table 4.** Sample C(100). Comparison of experimentally measured concentrations of analyzed volatile compounds in standard solutions obtained in NCRRIH&V-2 by three methods: cyclohexanol as IS, the ES method and method of using ethanol as IS with initial concentration according to the gravimetric method

| | acetaldehyde | methyl acetate | ethyl acetate | methanol | 2-propanol | ethanol | 1-propanol | isobutyl alcohol | isoamyl acetate | 1-butanol | isoamyl alcohol | ethyl hexanoate | cyclohexanol | ethyl octanoate | ethyl decanoate | benzyl alcohol | phenylethanol |
|---|---|---|---|---|---|---|---|---|---|---|---|---|---|---|---|---|---|
| | | | | | | **Gravimetric method** | | | | | | | | | | | |
| Concentration, mg /L (AA) | 98,1 | 94,2 | 94,8 | 123,7 | 93,9 | 789300 | 92,2 | 92,7 | 93,4 | 93,4 | 94,4 | 93,2 | 114,8 | 91,0 | 95,9 | 94,2 | 96,2 |
| amount, (0,5 mcl Split=20) pg | 2,3 | 2,3 | 2,3 | 3,0 | 2,2 | 18911 | 2,2 | 2,2 | 2,2 | 2,2 | 2,3 | 2,2 | 2,7 | 2,2 | 2,3 | 2,3 | 2,3 |
| response x10, pC | 5,5 | 4,4 | 6,3 | 7,1 | 7,8 | 61162 | 9,8 | 11,6 | 9,9 | 11,0 | 11,7 | 10,4 | 15,9 | 11,2 | 11,9 | 9,7 | 12,6 |
| | | | | | | **Cyclohesanol as IS** | | | | | | | | | | | |
| Concentration, mg /L (AA) | 94 | 88,02 | 89,02 | 124,70 | 88,14 | 799910 | 91,54 | 91,52 | 91,42 | 90,85 | 92,51 | 91,88 | 114,8 | 91,34 | 96,49 | 91,31 | 92,60 |
| repeat, % | 0,1 | 0,6 | 0,8 | 0,9 | 0,0 | 0,4 | 0,6 | 0,9 | 0,6 | 0,7 | 0,4 | 1,0 | 0,0 | 0,3 | 0,0 | 0,9 | 0,9 |
| relative bias, % | -4,1 | -6,6 | -6,1 | 0,8 | -6,1 | 1,3 | -0,7 | -1,3 | -2,1 | -2,7 | -2,0 | -1,4 | 0,0 | 0,4 | 0,6 | -3,1 | -3,7 |
| | | | | | | **ES** | | | | | | | | | | | |
| Concentration, mg /L (AA) | 110 | 102,6 | 103,9 | 145,4 | 101,2 | 801919 | 107,0 | 107,0 | 106,9 | 106,2 | 108,3 | 107,5 | 114,5 | 106,8 | 112,8 | 107,0 | 108,5 |
| repeat, % | 4,1 | 4,7 | 4,9 | 5,0 | 4,6 | 4,6 | 4,7 | 5,0 | 4,7 | 4,8 | 4,5 | 5,1 | 4,1 | 4,4 | 4,1 | 3,3 | 3,2 |
| relative bias, % | 12,1 | 8,9 | 9,6 | 17,6 | 7,7 | 1,6 | 16,1 | 15,4 | 14,5 | 13,7 | 14,7 | 15,3 | -0,3 | 17,4 | 17,6 | 13,6 | 12,8 |
| | | | | | | **Ethanol as IS** | | | | | | | | | | | |
| Concentration, mg /L (AA) | 93 | 87,10 | 88,09 | 123,39 | 87,22 | 789300 | 90,57 | 90,55 | 90,45 | 89,88 | 91,53 | 90,90 | 113,99 | 90,36 | 95,46 | 90,34 | 91,62 |
| repeat, % | 0,5 | 0,1 | 0,3 | 0,4 | 0,4 | 0,0 | 0,2 | 0,4 | 0,1 | 0,3 | 0,0 | 0,5 | 0,4 | 0,1 | 0,4 | 1,3 | 1,4 |
| relative bias, % | -5,2 | -7,5 | -7,1 | -0,3 | -7,1 | 0,0 | -1,8 | -2,3 | -3,2 | -3,8 | -3,0 | -2,5 | -0,7 | -0,7 | -0,5 | -4,1 | -4,8 |

**Table 5.** Sample D(10). Comparison of experimentally measured concentrations of analyzed volatile compounds in standard solutions obtained in NCRRIH&V-2 by three methods: cyclohexanol as IS, the ES method and method of using ethanol as IS with initial concentration according to the gravimetric method

| | acetaldehyde | methyl acetate | ethyl acetate | methanol | 2-propanol | ethanol | 1-propanol | isobutyl alcohol | isoamyl acetate | 1-butanol | isoamyl alcohol | ethyl hexanoate | cyclohexanol | ethyl octanoate | ethyl decanoate | benzyl alcohol | phenylethanol |
|---|---|---|---|---|---|---|---|---|---|---|---|---|---|---|---|---|---|
| | | | | | | | | | | **Gravimetric method** | | | | | | | |
| Concentration, mg /L (AA) | 8,06 | 7,65 | 7,71 | 33,66 | 7,72 | 789300 | 7,49 | 7,54 | 7,59 | 7,59 | 7,67 | 7,57 | 93,51 | 7,40 | 7,79 | 7,65 | 7,82 |
| amount, (0,5 mcl Split=20) pg | 0,193 | 0,184 | 0,185 | 0,808 | 0,185 | 18938 | 0,180 | 0,181 | 0,182 | 0,182 | 0,184 | 0,182 | 2,244 | 0,178 | 0,187 | 0,184 | 0,188 |
| response x10, pC | 0,49 | 0,41 | 0,56 | 1,93 | 0,66 | 56909 | 0,85 | 0,99 | 0,87 | 0,94 | 1,02 | 0,89 | 12,73 | 0,95 | 1,00 | 0,85 | 1,12 |
| | | | | | | | | | | **Cyclohesanol as IS** | | | | | | | |
| Concentration, mg /L (AA) | 8,56 | 8,27 | 8,06 | 34,55 | 7,70 | 765233 | 8,04 | 8,07 | 8,21 | 8,23 | 8,22 | 8,11 | 93,51 | 8,03 | 8,41 | 8,19 | 8,48 |
| repeat, % | 3,0 | 4,5 | 1,3 | 0,3 | 2,1 | 1,7 | 2,9 | 2,7 | 4,4 | 0,9 | 0,1 | 2,2 | 0,0 | 1,4 | 4,7 | 0,9 | 3,1 |
| relative bias, % | 2,2 | 3,9 | 0,7 | 1,7 | -4,0 | -3,0 | 3,3 | 3,0 | 4,1 | 4,4 | 3,1 | 3,1 | 0,0 | 4,4 | 3,9 | 3,0 | 4,3 |
| | | | | | | | | | | **ES** | | | | | | | |
| Concentration, mg /L (AA) | 10,55 | 9,69 | 9,46 | 40,54 | 9,08 | 773603 | 9,46 | 9,75 | 9,67 | 9,68 | 9,68 | 9,55 | 93,78 | 9,45 | 10,00 | 9,66 | 9,99 |
| repeat, % | 9,0 | 3,5 | 0,3 | 0,8 | 3,1 | 2,7 | 1,9 | 3,5 | 5,5 | 1,9 | 0,9 | 3,2 | 1,0 | 2,4 | 3,5 | 1,9 | 4,1 |
| relative bias, % | 25,9 | 21,9 | 18,2 | 19,3 | 13,1 | -2,0 | 21,5 | 24,5 | 22,5 | 22,7 | 21,3 | 21,3 | 0,3 | 22,9 | 23,5 | 21,4 | 22,9 |
| | | | | | | | | | | **Ethanol as IS** | | | | | | | |
| Concentration, mg /L (AA) | 8,83 | 8,53 | 8,32 | 35,66 | 7,94 | 789300 | 8,30 | 8,44 | 8,47 | 8,50 | 8,48 | 8,37 | 95,69 | 8,28 | 8,68 | 8,45 | 8,74 |
| repeat, % | 1,3 | 6,2 | 3,0 | 1,9 | 0,4 | 0,0 | 4,5 | 3,6 | 2,8 | 0,8 | 1,8 | 0,5 | 1,7 | 0,3 | 3,1 | 0,7 | 1,4 |
| relative bias, % | 5,4 | 7,3 | 3,9 | 5,0 | -1,0 | 0,0 | 6,6 | 7,7 | 7,4 | 7,7 | 6,3 | 6,4 | 2,3 | 7,7 | 7,1 | 6,2 | 7,6 |

**Table 6.** Sample E(2). Comparison of experimentally measured concentrations of analyzed volatile compounds in standard solutions obtained in NCRRIH&V-2 by three methods: cyclohexanol as IS, the ES method and method of using ethanol as IS with initial concentration according to the gravimetric method

| | acetaldehyde | methyl acetate | ethyl acetate | methanol | 2-propanol | ethanol | 1-propanol | isobutyl alcohol | isoamyl acetate | 1-butanol | isoamyl alcohol | ethyl hexanoate | cyclohexanol | ethyl octanoate | ethyl decanoate | benzyl alcohol | phenylethanol |
|---|---|---|---|---|---|---|---|---|---|---|---|---|---|---|---|---|---|
| | | | | | | | | | **Gravimetric method** | | | | | | | | |
| Concentration, mg /L (AA) | 1,871 | 1,702 | 1,714 | 27,471 | 1,795 | 789300 | 1,666 | 1,676 | 1,689 | 1,689 | 1,707 | 1,684 | 92,139 | 1,646 | 1,733 | 1,702 | 1,740 |
| amount, (0,5 mcl Split=20) pg | 0,045 | 0,041 | 0,041 | 0,659 | 0,043 | 18938 | 0,040 | 0,040 | 0,041 | 0,041 | 0,041 | 0,040 | 2,211 | 0,039 | 0,042 | 0,041 | 0,042 |
| response x10, pC | 0,104 | 0,083 | 0,107 | 1,476 | 0,135 | 55264 | 0,156 | 0,194 | 0,171 | 0,182 | 0,182 | 0,167 | 11,433 | 0,180 | 0,216 | 0,164 | 0,228 |
| | | | | | | | | | **Cyclohesanol as IS** | | | | | | | | |
| Concentration, mg /L (AA) | 2,00 | 1,80 | 1,74 | 28,3 | 1,77 | 802125 | 1,69 | 1,68 | 1,71 | 1,74 | 1,68 | 1,72 | 92,10 | 1,61 | 1,89 | 1,72 | 1,86 |
| repeat, % | 2,5 | 9,0 | 4,0 | 5,2 | 7,3 | 1,6 | 6,5 | 4,4 | 1,1 | 2,8 | 7,5 | 7,3 | 0,0 | 4,5 | 6,1 | 0,6 | 0,2 |
| relative bias, % | 3,3 | 1,7 | -2,4 | 2,9 | -5,0 | 1,6 | -2,6 | -3,8 | -2,3 | -0,8 | -5,5 | -1,8 | 0,0 | -5,6 | 4,9 | -2,6 | 2,9 |
| | | | | | | | | | **ES** | | | | | | | | |
| Concentration, mg /L (AA) | 2,31 | 1,95 | 1,76 | 30,1 | 1,61 | 734643 | 1,62 | 1,79 | 1,83 | 1,86 | 1,79 | 1,83 | 83,69 | 1,83 | 2,01 | 1,84 | 1,99 |
| repeat, % | 1,3 | 12,7 | 1,1 | 5,4 | 1,1 | 1,8 | 13,1 | 4,6 | 0,9 | 13,1 | 7,3 | 7,1 | 0,2 | 16,2 | 6,3 | 0,4 | 0,0 |
| relative bias, % | 19,3 | 10,1 | -1,3 | 9,3 | -13,7 | -6,9 | -6,6 | 2,5 | 4,1 | 5,9 | 0,8 | 4,6 | -9,1 | 6,9 | 11,8 | 4,1 | 9,9 |
| | | | | | | | | | **Ethanol as IS** | | | | | | | | |
| Concentration, mg /L (AA) | 1,97 | 1,77 | 1,71 | 27,9 | 1,74 | 789300 | 1,70 | 1,65 | 1,69 | 1,71 | 1,65 | 1,69 | 89,91 | 1,59 | 1,86 | 1,70 | 1,83 |
| repeat, % | 4,1 | 7,5 | 5,5 | 3,6 | 8,8 | 0,0 | 12,3 | 2,9 | 2,7 | 1,3 | 9,1 | 8,9 | 1,6 | 2,9 | 4,5 | 2,2 | 1,7 |
| relative bias, % | 1,7 | 0,1 | -3,9 | 1,2 | -6,5 | 0,0 | -1,9 | -5,3 | -3,9 | -2,3 | -6,9 | -3,3 | -2,4 | -7,1 | 3,3 | -4,1 | 1,3 |

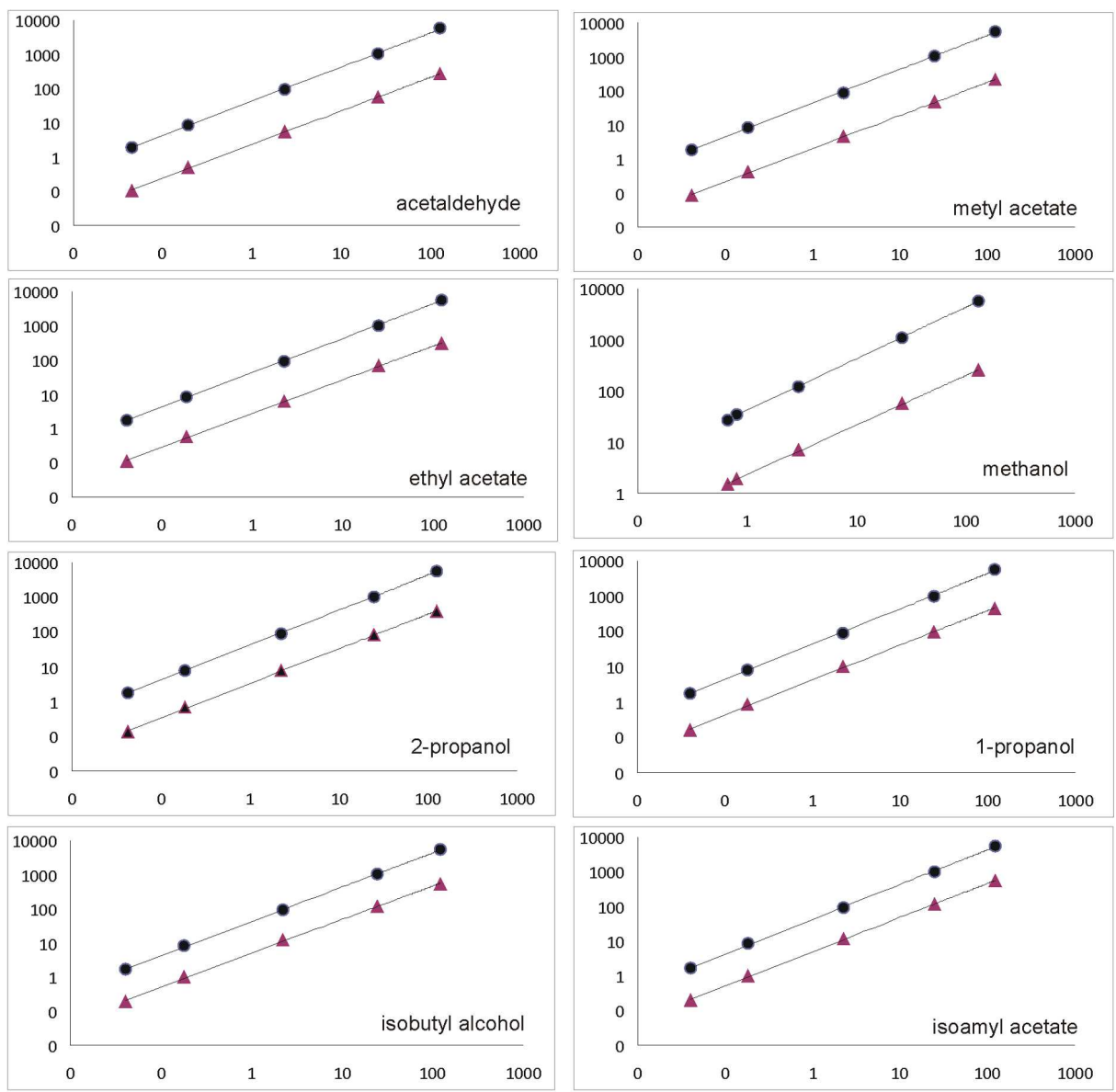

**Fig. 1.** Experimental results from NCRRIH&V-2 for the following compounds: acetaldehyde, methyl acetate, ethyl acetate, methanol, 2-propanol, 1-propanol, isobutyl alcohol, isoamyl acetate. ● – concentration, mg/L (AA), ▲ – response×10, pC, horizontal axis – amount, pg.

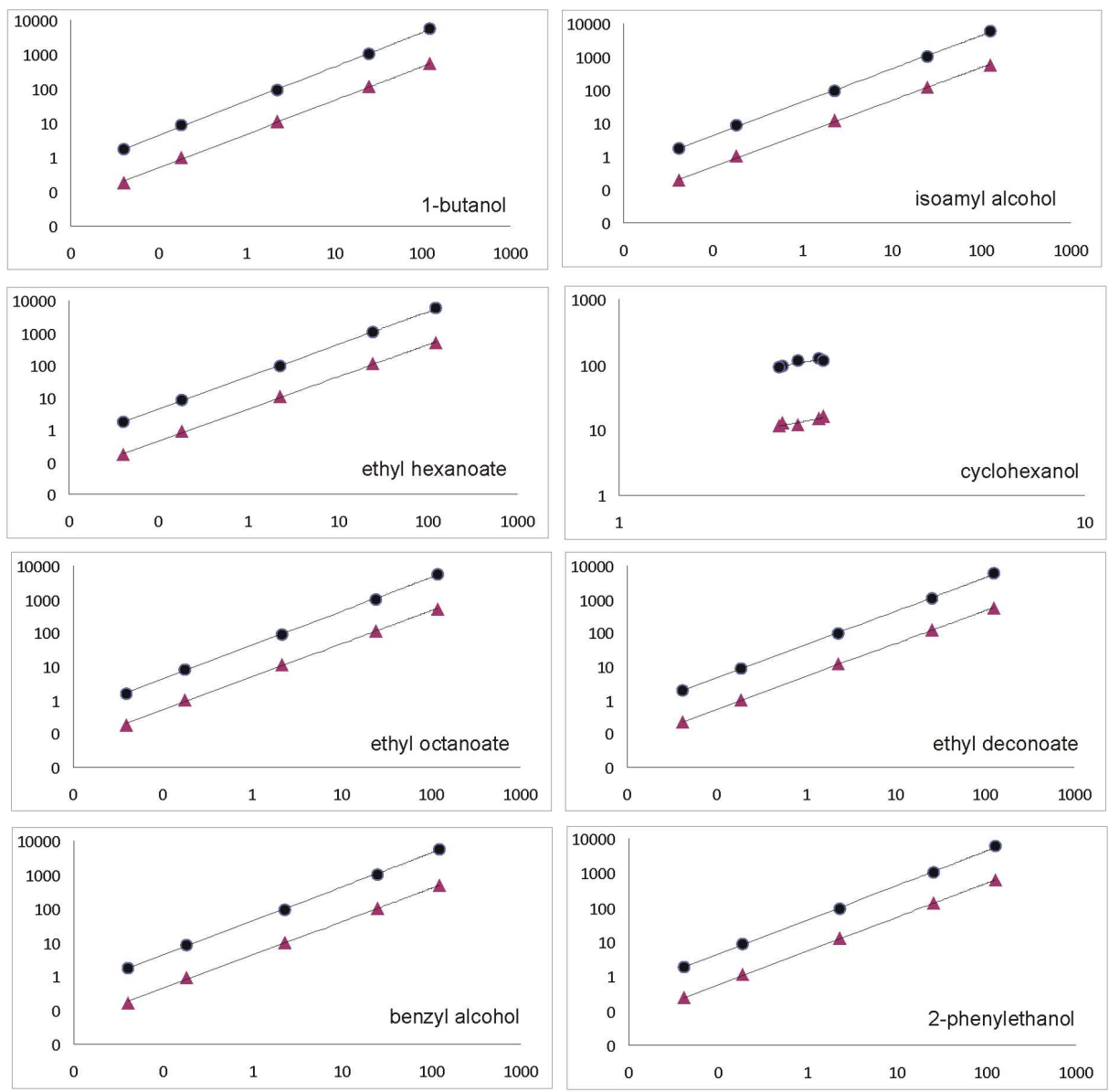

**Fig 2.** Experimental results from NCRRIH&V-2 for the following compounds: 1-butanol, isoamyl alcohol, ethyl hexanoate, cyclohexanol, ethyl octanoate, ethyl decanoate, benzyl alcohol and 2-phenylethanol. ● – concentration, mg/L (AA), ▲ – response×10, pC, horizontal axis – amount, pg.

Concentrations of volatile compounds in these solutions A(5000), B(1000), C(100), D(10) and E(2) in accordance with gravimetric method and calculated concentrations on the base of measured raw data are given in Tables 2–6. These solutions were examined by three methods: IS method, ES method and "Ethanol as IS". For the first method cyclohexanol was added as IS.

For illustrative purposes the experimental data are presented in Figs. 1, 2 for the following main analyzed compounds: acetaldehyde, methyl acetate, ethyl acetate, methanol, 2-propanol, 1-propanol, isobutyl alcohol, isoamyl acetate (Fig. 1), 1-butanol, isoamyl alcohol, ethyl hexanoate, cyclohexanol, ethyl octanoate, ethyl decanoate, benzyl alcohol and 2-phenylethanol (Fig. 2). The presented graphs show the linear dependence of the detector response (triangle marked) and concentration (circle marked) in mg/L of AA on the amount of the examined component coming directly to the detector. Fig. 3 contains chromatograms for these above-named compounds and ethanol for solutions A(5000), B(1000), C(100), D(10) and E(2).

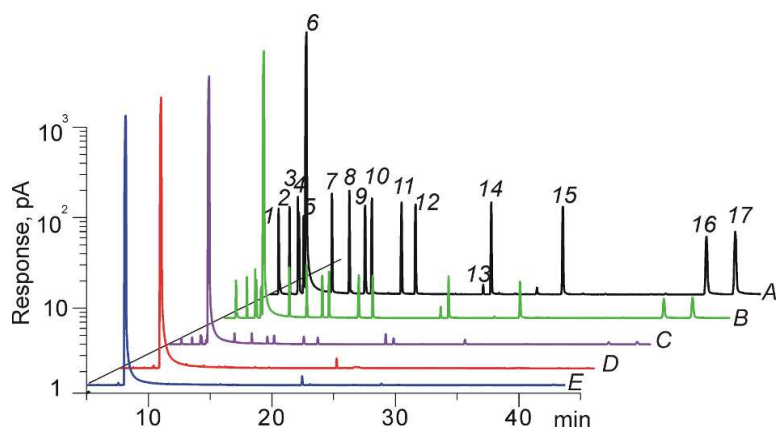

**Fig. 3.** Chromatograms of standard solutions A–E from Tables 2–6. 1 – acetaldehyde, 2 – methyl acetate, 3 – ethyl acetate, 4 – methanol, 5 – 2-propanol, 6 – ethanol, 7 – 1-propanol, 8 – isobutyl alcohol, 9 – isoamyl acetate, 10 – 1-butanol, 11 – isoamyl alcohol, 12 – ethyl hexanoate, 13 – cyclohexanol, 14 – ethyl octanoate, 15 – ethyl decanoate, 16 – benzyl alcohol, 17 – 2-phenylethanol.

The analysis of the experimental data shows that the relative bias between the experimentally measured concentrations calculated in accordance with proposed method using ethanol as IS and the values of concentrations assigned during the preparation by gravimetric method for all analyzed fifteen components in the five analyzed working solutions do not exceed 7,7 %. At the same time the relative bias between measured concentrations calculated in accordance with traditional methods IS and ES with values of concentrations assigned during the preparation by gravimetric method for all analyzed fifteen components in the five analyzed solutions does not exceed 6,6 % (IS method using cyclohexanol as IS) and 25,9 % ( ES method).

*Experiments with ethanol-water solutions like vodka and whisky*

The second series of experiments have been carried out in laboratories INP BSU and CLCheb on the GC Chromatec-Crystal5000 and HP6890, respectively. To demonstrate the reliability of the proposed method, the standard ethanol-water (96:4) solution with initial volatile compounds concentration about 4000 mg/L of AA was analyzed after dilution with water in the ratios 1:1, 1:9, 1:99, 1:1999 and 1:9999. Experimental results are presented in Table 7 (INP BSU results) and Table 8 (CLCheb results). Illustrations of obtained experimental data are given in Figs. 4, 5 (INP BSU results) and Figs. 6, 7 (CLCheb results). In Figs. 4, 6 the circle marked line is concentration of the analysed compound expressed in mg per litre of absolute alcohol. The second line (triangle marked) and the square marked one are the detector response versus the amount of the compound and the concentration in mg per 1 L of solution, respectively. Figs. 4, 5 contain data for the following compounds: acetaldehyde, methyl acetate, ethyl acetate, methanol, 2-propanol, ethanol, 1-propanol, isobutyl alcohol, 1-butanol, isoamyl alcohol. Fig. 5 presents chromatograms for these compounds for standard working solutions A (No dilution), B (1:1), C (1:9), D (1:99), E (1:999), F (1:9999). Figs. 6 and 7 demonstrates data for acetaldehyde, methyl acetate, methanol, 2-propanol, ethanol, 1-propanol, isobutyl alcohol, 1-butanol for standard working solutions A (no dilution), B (1:1), C (1:9), D (1:99), E (1:999).

**Table 7.** Concentrations of analyzed volatile compounds and ethanol, presented according to water dilution (INP BSU)

| sample (dilution) | | | | | | | | | | |
|---|---|---|---|---|---|---|---|---|---|---|
| | concentration under gravimetric method, mg /L (AA) | | | | | | | | | |
| | measured concentration, mg /L (AA) | | | | | | | | | |
| | relative bias,% | | | | | | | | | |
| | concentration under gravimetric method, mg /L (sol) | | | | | | | | | |
| | amount, pg | | | | | | | | | |
| | response×10, pC | | | | | | | | | |
| | acetaldehyde | methyl acetate | ethyl acetate | methanol | 2-propanol | ethanol | 1-propanol | isobutyl alcohol | 1-butanol | isoamyl alcohol |
| A (No) | 4275 | 4397 | 4173 | 41995 | 3991 | 789300 | 4012 | 3975 | 4071 | 4071 |
| | 4556 | 4436 | 4253 | 42586 | 4112 | N/A | 4076 | 4049 | 4174 | 4458 |
| | 6,6 | 0,9 | 1,9 | 1,4 | 3,0 | N/A | 1,6 | 1,9 | 2,5 | 9,5 |
| | 3768 | 3875 | 3678 | 37017 | 3518 | 695748 | 3536 | 3504 | 3588 | 3588 |
| | 91899 | 94524 | 89710 | 902864 | 85806 | 16969460 | 86253 | 85466 | 87522 | 87522 |
| | 10720 | 11276 | 14420 | 115328 | 16655 | 2825852 | 19676 | 22784 | 21330 | 23143 |
| B (1:1) | 4275 | 4397 | 4173 | 41995 | 3991 | 789300 | 4012 | 3975 | 4071 | 4071 |
| | 4451 | 4127 | 4018 | 40462 | 4000 | N/A | 3973 | 4007 | 4096 | 4412 |
| | 4,1 | -6,1 | -3,7 | -3,7 | 0,2 | N/A | -1,0 | 0,8 | 0,6 | 8,4 |
| | 1884 | 1938 | 1839 | 18509 | 1759 | 347874 | 1768 | 1752 | 1794 | 1794 |
| | 43761 | 45012 | 42719 | 429935 | 40860 | 8080695 | 41073 | 40698 | 41677 | 41677 |
| | 4733 | 4741 | 6157 | 49525 | 7323 | 1277251 | 8668 | 10190 | 9462 | 10353 |
| C (1:9) | 4275 | 4397 | 4173 | 41995 | 3991 | 789300 | 4012 | 3975 | 4071 | 4071 |
| | 4340 | 3961 | 3780 | 39043 | 3875 | N/A | 3868 | 3904 | 4012 | 4318 |
| | 1,5 | -9,9 | -9,4 | -7,0 | -2,9 | N/A | -3,6 | -1,8 | -1,4 | 6,1 |
| | 377 | 388 | 368 | 3702 | 352 | 69575 | 354 | 350 | 359 | 359 |
| | 9190 | 9452 | 8971 | 90286 | 8581 | 1696946 | 8625 | 8547 | 8752 | 8752 |
| | 931,6 | 918,7 | 1169 | 9647 | 1432 | 257842 | 1704 | 2004 | 1870 | 2045 |
| D (1:99) | 4275 | 4397 | 4173 | 41995 | 3991 | 789300 | 4012 | 3975 | 4071 | 4071 |
| | 4406 | 4002 | 3762 | 38645 | 3866 | N/A | 3862 | 3903 | 4107 | 4479 |
| | 3,1 | -9,0 | -9,8 | -8,0 | -3,1 | N/A | -3,7 | -1,8 | 0,9 | 10,0 |
| | 37,7 | 38,8 | 36,8 | 370,2 | 35,2 | 6958 | 35,4 | 35,0 | 35,9 | 35,9 |
| | 919 | 945 | 897 | 9029 | 858 | 169695 | 863 | 855 | 875 | 875 |
| | 78,57 | 77,10 | 96,66 | 793,5 | 118,8 | 21427 | 141,33 | 166,6 | 159,1 | 171,9 |
| E (1:999) | 4275 | 4397 | 4173 | 41995 | 3991 | 789300 | 4012 | 3975 | 4071 | 4071 |
| | 4280 | 4292 | 4107 | 38764 | 3818 | N/A | 3820 | 4140 | 4024 | 3937 |
| | 0,1 | -2,4 | -1,6 | -7,7 | -4,3 | N/A | -4,8 | 4,1 | -1,2 | -3,3 |
| | 3,77 | 3,88 | 3,68 | 37,02 | 3,52 | 696 | 3,54 | 3,50 | 3,59 | 3,59 |
| | 91,9 | 94,5 | 89,7 | 903 | 85,8 | 16969 | 86,3 | 85,5 | 87,5 | 87,5 |
| | 7,74 | 8,60 | 10,7 | 80,7 | 12,4 | 2173 | 14,2 | 17,9 | 15,8 | 15,8 |
| F (1:9999) | N/A | N/A | N/A | 41995 | N/A | 789300 | N/A | N/A | N/A | N/A |
| | | | | 39210 | | N/A | | | | |
| | | | | -6,6 | | N/A | | | | |
| | | | | 3,702 | | 69,6 | | | | |
| | | | | 90,3 | | 1697 | | | | |
| | | | | 8,12 | | 223 | | | | |

**Table 8.** Concentrations of analyzed volatile compounds and ethanol, presented according to water dilution (CLCheb)

| sample (dilution) | | | | concentration under gravimetric method, mg /L (AA) measured concentration, mg /L (AA) relative bias,% concentration under gravimetric method, mg /L (sol) amount, ng response×10, pC | | | | |
|---|---|---|---|---|---|---|---|---|
| | acetaldehyde | methyl acetate | methanol | 2-propanol | ethanol | 1-propanol | isobutyl alcohol | 1-butanol |
| A (No) | 5170 | 5200 | 5291 | 5242 | 789300 | 7196 | 1171 | 5324 |
| | 5405 | 5127 | 4840 | 5091 | 789300 | 7157 | 1157 | 5358 |
| | 4,5 | -1,4 | -8,5 | -2,9 | N/A | -0,5 | -1,2 | 0,6 |
| | 4756 | 4784 | 4868 | 4823 | 726156 | 6620 | 1077 | 4898 |
| | 119 | 120 | 122 | 121 | 18154 | 166 | 27 | 122 |
| | 262640 | 473048 | 551182 | 785344 | 117394309 | 1038326 | 289264 | 1635734 |
| B (1:1) | 5170 | 5200 | 5291 | 5242 | 789300 | 7196 | 1171 | 5324 |
| | 5231 | 5292 | 5288 | 5243 | 789300 | 7187 | 1174 | 5311 |
| | 1,2 | 1,8 | -0,1 | 0,0 | N/A | -0,1 | 0,2 | -0,3 |
| | 2378 | 2392 | 2434 | 2412 | 363078 | 3310 | 539 | 2449 |
| | 59,5 | 59,8 | 60,8 | 60,3 | 9077,0 | 82,8 | 13,5 | 61,2 |
| | 131421 | 256031 | 309192 | 422395 | 60259800 | 534562 | 150959 | 831805 |
| C (1:9) | 5170 | 5200 | 5291 | 5242 | 789300 | 7196 | 1171 | 5324 |
| | 3936 | 6849 | 5107 | 9092 | 789300 | 5219 | 1216 | 6861 |
| | -1,0 | -1,8 | 0,8 | -1,3 | N/A | -0,5 | -1,5 | -0,8 |
| | 476 | 478 | 487 | 482 | 72616 | 662 | 108 | 490 |
| | 11,9 | 12,0 | 12,2 | 12,1 | 1815,4 | 16,6 | 2,7 | 12,2 |
| | 28419 | 55123 | 70780 | 89072 | 12827790 | 116146 | 31552 | 176935 |
| D (1:99) | 5170 | 5200 | 5291 | 5242 | 789300 | 7196 | 1171 | 5324 |
| | 3815 | 6781 | 4919 | 8321 | 789300 | 5075 | 1167 | 6648 |
| | 0,5 | 1,3 | 0,0 | -0,2 | N/A | 0,1 | 0,0 | -0,2 |
| | 47,6 | 47,8 | 48,7 | 48,2 | 7261,6 | 66,2 | 10,8 | 49,0 |
| | 1,19 | 1,20 | 1,22 | 1,21 | 181,54 | 1,66 | 0,27 | 1,22 |
| | 3267 | 6580 | 7903 | 9631 | 1499539 | 13284 | 3594 | 20158 |
| E (1:999) | 5170 | 5200 | 5291 | 5242 | 789300 | 7196 | 1171 | 5324 |
| | 4474 | 5388 | 5233 | 4717 | 789300 | 6553 | 1090 | 4803 |
| | -13,5 | 3,6 | -1,1 | -10,0 | N/A | -8,9 | -7,0 | -9,8 |
| | 4,76 | 4,79 | 4,87 | 4,83 | 726,88 | 6,63 | 1,08 | 4,90 |
| | 0,119 | 0,120 | 0,122 | 0,121 | 18,172 | 0,166 | 0,027 | 0,123 |
| | 0,119 | 0,120 | 0,122 | 0,121 | 18,172 | 0,166 | 0,027 | 0,123 |

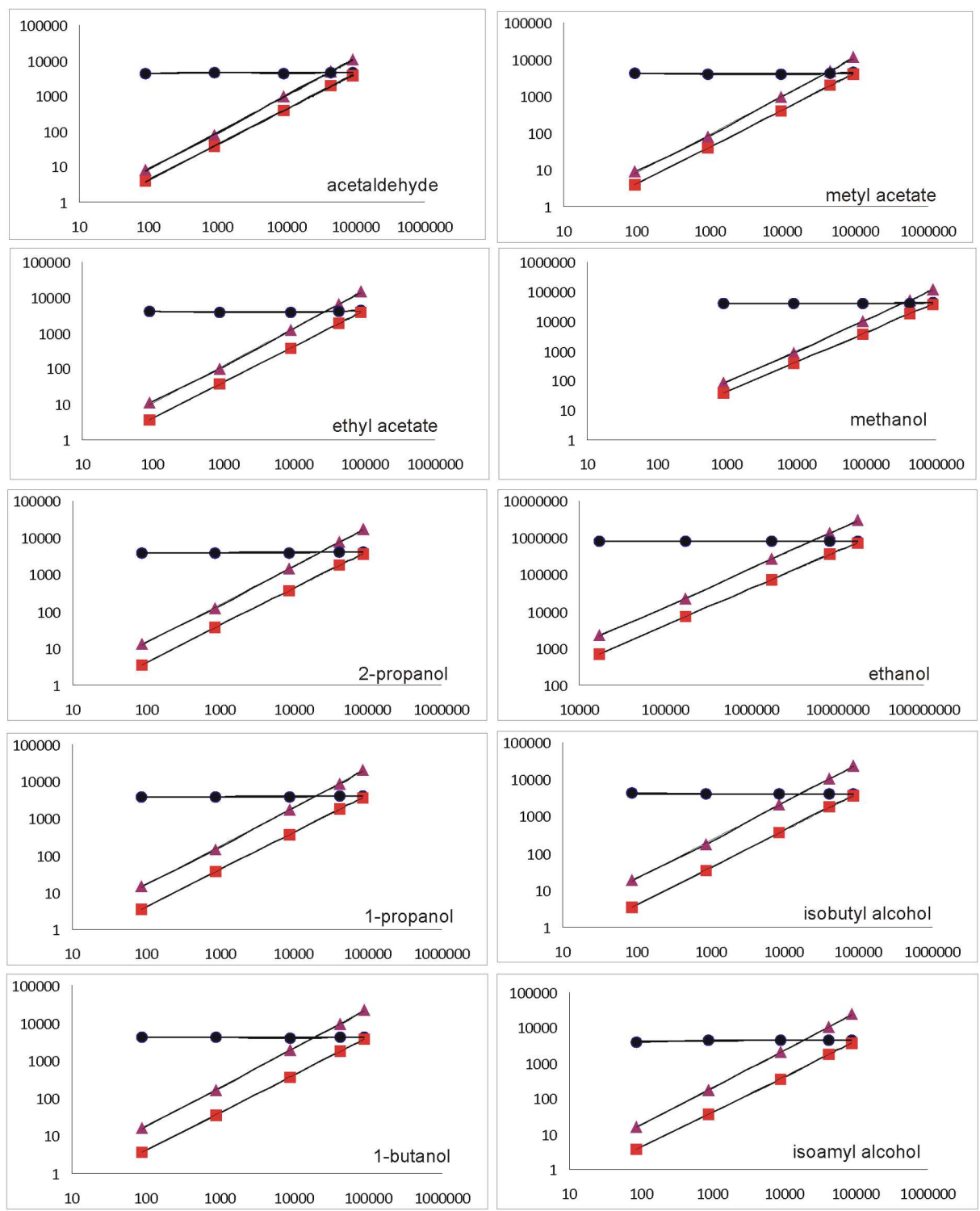

**Fig. 4.** Experimental results (INP BSU) for the following compounds: acetaldehyde, methyl acetate, ethyl acetate, methanol, 2-propanol, ethanol, 1-propanol, isobutyl alcohol, 1-butanol, isoamyl alcohol. ● – concentration, mg/L (AA), ■ – concentration, mg/L (sol), ▲ – response×10, pC, horizontal axis – amount, pg.

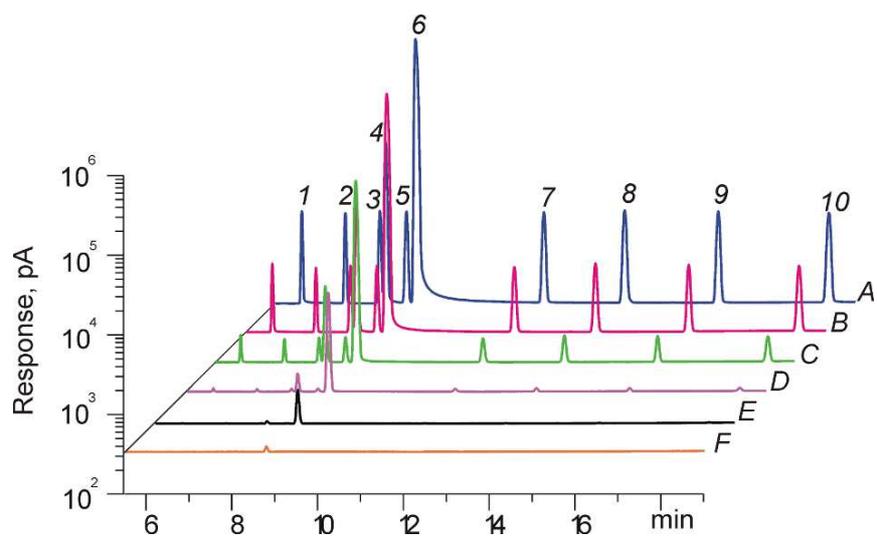

**Fig. 5.** Chromatograms (INP BSU) of standard solutions A–F corresponding to Table 7. 1 – acetaldehyde, 2 – methyl acetate, 3 – ethyl acetate, 4 – methanol, 5 – 2-propanol, 6 – ethanol, 7 – 1-propanol, 8 – isobutyl alcohol, 9 – 1-butanol, 10 – isoamyl alcohol.

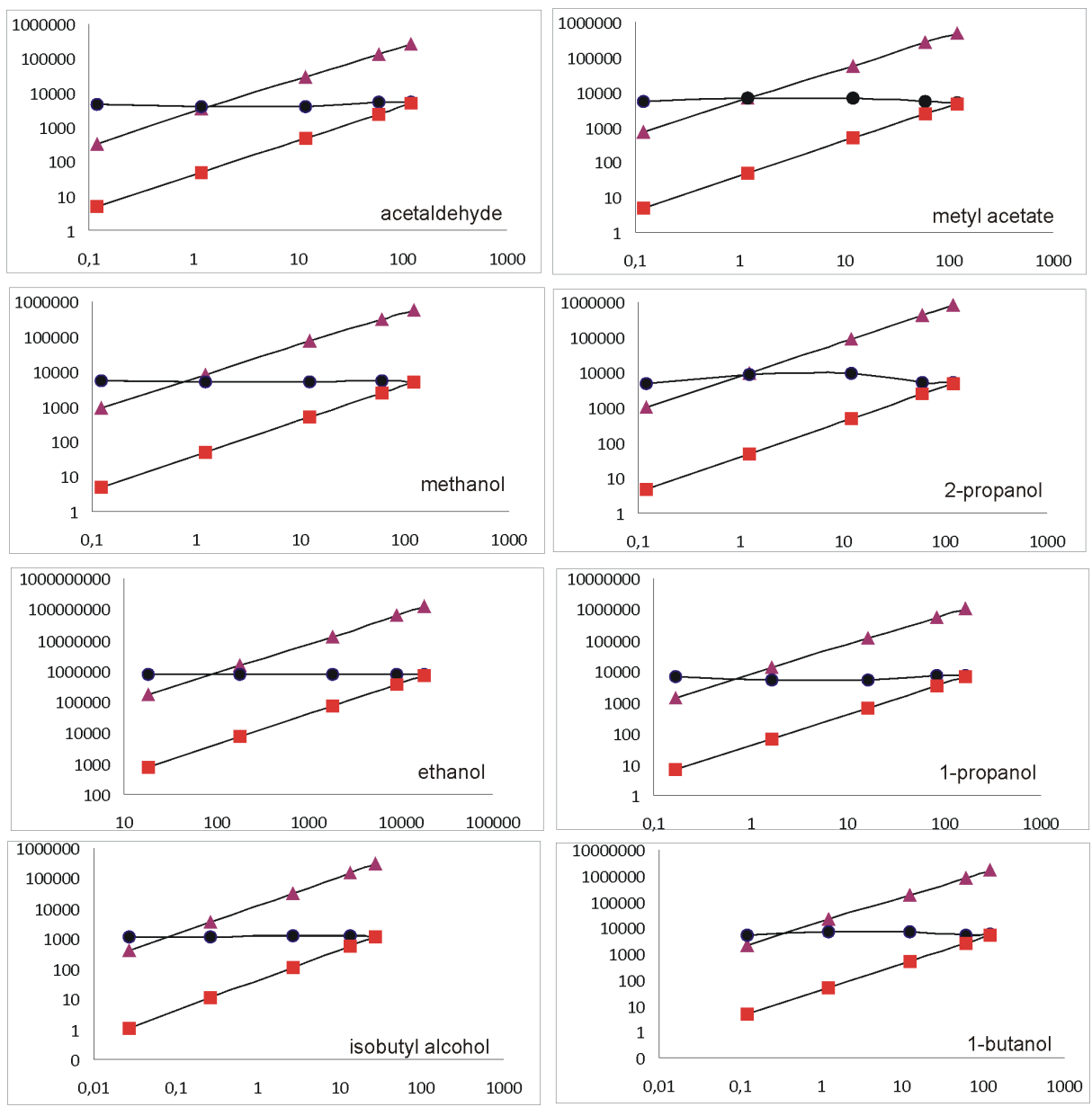

Fig. 6. Experimental results (CLCheb) for the following compounds: acetaldehyde, methyl acetate, methanol, 2-propanol, ethanol, 1-propanol, isobutyl alcohol, 1-butanol. ● – concentration, mg/L (AA), ■ – concentration, mg/L (sol), ▲ – response×10, pC, horizontal axis – amount, pg.

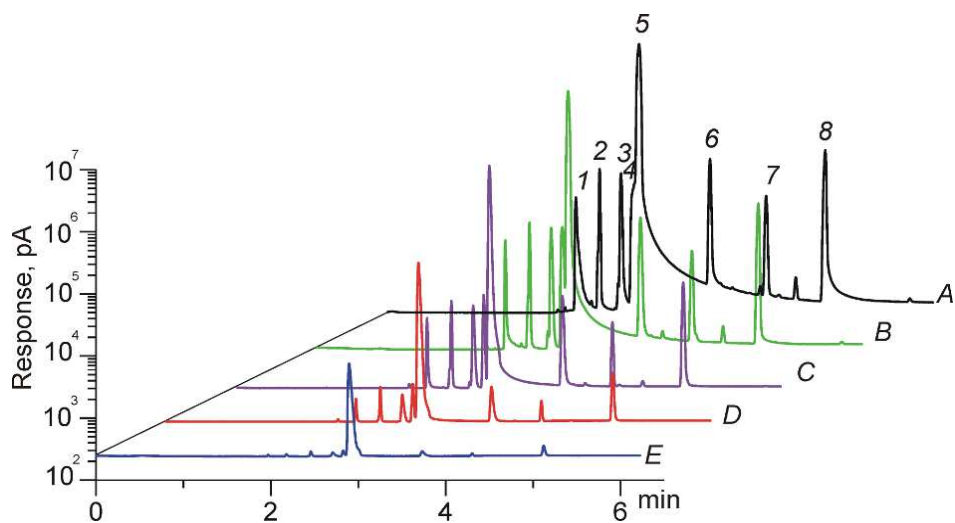

Fig. 7. Chromatograms (CLCheb) of standard solutions A–E from Table 8. 1 – acetaldehyde, 2 – methyl acetate, 3 – methanol, 4 – 2-propanol, 5 – ethanol, 6 – 1-propanol, 7 – isobutyl alcohol, 8 – 1-butanol.

Even after dilution of the initial solution with water in the ratio 1:999, the difference between the measured concentrations of all compounds and their values calculated using the gravimetric method does not exceed 7.8 %. In the sample with dilution 1:9999 there are only peaks of methanol and ethanol. Other compounds are significantly less than the level of detection. But even in this case the relative discrepancy of measured concentrations of methanol does not exceed 6.6%.

*Experiments with certified reference material CRM LGC5100 Whisky-Congeners*

For demonstration of robustness of the new method for direct determination of volatile compounds in spirit products the results of analysis of certified reference material CRM LGC5100 Whisky-Congeners are presented. Certified values of analyzed volatile compound concentrations and their uncertainties are given in Table 9. The measured chromatograms are shown in Figs. 8–10. Experiments were carried out in INP BSU, NCRRIH&V-1 and NCRRIH&V-2.

In all three laboratories examination was carried out according to the new method using ethanol as IS. In Fig. 8 fragment of chromatograms from INP BSU with interesting peaks is presented in logarithmic scale and linear scale (in the right upper corner). The following peaks are indicated here: methanol (1), ethanol (2), 1-propanol (3), isobutanol (4),

1-butanol (5) and isoamilol (6). Fig. 9 and 10 contain the analogous data from NCRRIH&V-1 and NCRRIH&V-2, respectively.

The determined values of concentrations of analyzed compounds are presented in Table 9 along with data of certified reference material CRM LGC5100 Whisky-Congeners. In this Table one can see certified value in g/100 L (AA),A) and relative uncertainty in % for certified reference material under certificate; values of RRF, peak area (a.u.), calculated concentration in g/100 L (AA) and relative bias in % for experimental data. Rightmost column contains values of RRF averaged between three laboratories. Calculation shows that the relative standard deviation of the values of RRF for all analyzing compounds does not exceed 6.1%. These predetermined values of RRF can be used in other laboratories for other chromatographs for evaluated measurements. Subsequent refinement of the values RRF can be performed individually for each GC with help of standard solutions in accordance with (4).

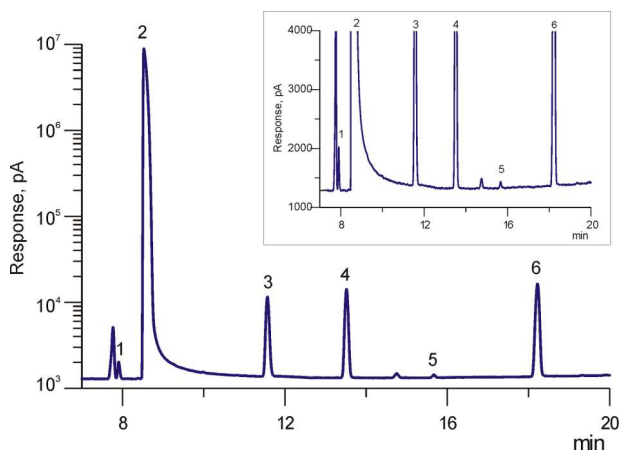

Fig. 8. Fragment of chromatogram from INP BSU with following peaks: 1 – methanol, 2 – ethanol, 3 – 1-propanol, 4 – isobutanol, 5 – 1-butanol, 6 – isoamilol.

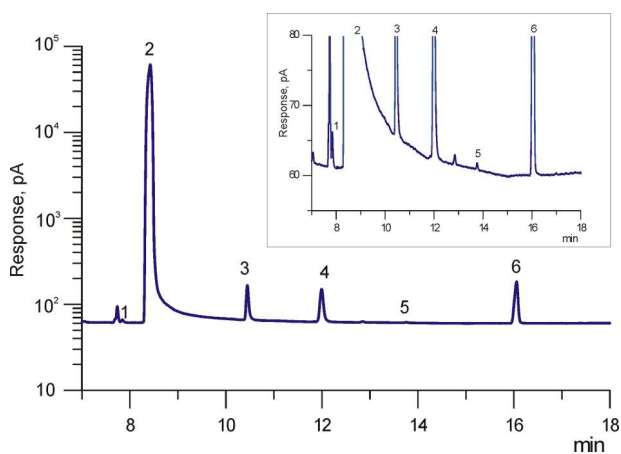

Fig. 9. Fragment of chromatogram from NCRRIH&V-1. 1 – methanol, 2 – ethanol, 3 – 1-propanol, 4 – isobutanol, 5 – 1-butanol, 6 – isoamilol.

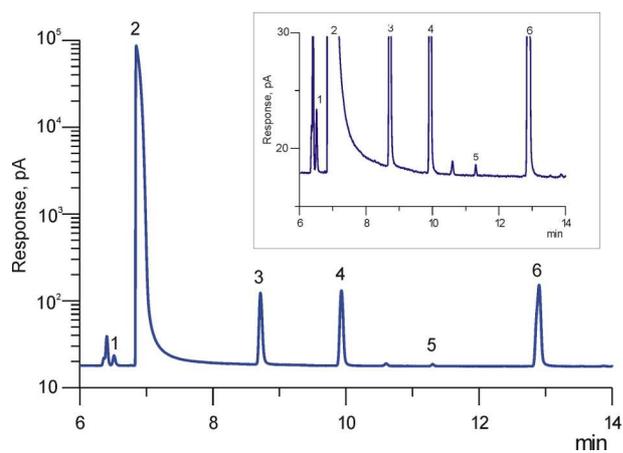

Fig. 10. Fragment of chromatogram from NCRRIH&V-2. 1 – methanol, 2 – ethanol, 3 – 1-propanol, 4 – isobutanol, 5 – 1-butanol, 6 – isoamilol.

In the certification procedure of CRM LGC5100 Whisky-Congeners as the inter-laboratory study sixteen profile authority laboratories took part. We have to conclude that our results of investigation this reference material are in good agreement with certified values as one can see from Table 9.

**Table 9.** Values of concentrations of analyzed compounds obtained from laboratories INP BSU, NCRRIH&V-1 and NCRRIH&V-2

| CRM LGC 5100 Whisky – Congeners Data of issue: November 2011 | | | INP BSU | | | | NCRRIH&V-1 | | | | NCRRIH&V-2 | | | | RRF (aver.) |
|---|---|---|---|---|---|---|---|---|---|---|---|---|---|---|---|
| Compound | Certified value, g/100 L (AA) | Uncertainty, % | Peak area, a.u. | RRF | C, g/100 L (AA) | Relative bias, % | Peak area, a.u. | RRF | C, g/100 L (AA) | Relative bias, % | Peak area, a.u. | RRF | C, g/100 L (AA) | Relative bias, % | |
| methanol | 5,2 | 6,2 | 0,3959 | 1,217 | 5,25 | 1,0 | 0,00249 | 1,361 | 4,91 | -6,5 | 0,002347 | 1,347 | 4,83 | -1,7 | 1,308 |
| ethanol | | | 7238,9 | 1,000 | | | 54,51 | 1,000 | | | 51,66 | 1,000 | | | 1,000 |
| 1-propanol | 57 | 4,2 | 7,856 | 0,676 | 57,91 | 5,8 | 0,05355 | 0,719 | 55,75 | -3,7 | 0,04456 | 0,717 | 48,81 | -12,4 | 0,704 |
| isobutanol | 58,8 | 5,3 | 9,76 | 0,579 | 61,62 | 8,1 | 0,06547 | 0,616 | 58,40 | -5,2 | 0,05351 | 0,604 | 49,38 | -15,4 | 0,600 |
| 1-butanol | 0,48 | 22,9 | 0,074 | 0,645 | 0,52 | 15,4 | 0,00053 | 0,668 | 0,51 | -1,5 | 0,0004312 | 0,64 | 0,42 | -17,8 | 0,651 |
| isoamylol | 79,58 | 3,5 | 12,931 | 0,63 | 88,83 | 10,8 | 0,08768 | 0,639 | 81,13 | -8,7 | 0,07437 | 0,607 | 68,97 | -15,0 | 0,625 |

## Supplementary data

The proposed method can be applied in other areas of food control. For example, there is the following international standard ISO 10315 "Cigarettes. Determination of nicotine in smoke condensates. Gas Chromatographic method".

In accordance with ISO 10315 five conditioned cigarettes are smoked using a 20 port linear smoking machine. The mainstream smoke is collected on a 44 mm Cambridge filter pad (CFP). After smoking, the CFP is extracted with propan-2-ol (isopropyl alcohol – IPA) containing heptadecane as internal standard. (IS) In accordance with proposed method the main component of solution IPA can be used as IS. Corresponding experimental results obtained in the laboratory INP BSU are given in the Table 10.

There were prepared nine standard working solutions of nicotine in IPA by gravimetric method. In accordance with ISO 10315 heptadecane was added as IS. Obtained concentrations are given in the second column of Table 10. Experimentally measured data were calculated by two methods – main component (IPA) as IS and heptadecane as IS. One can see good agreement between results by these methods. This is illustrated in Fig. 11.

**Table 10.** Experiments with standard working solutions of nicotine in IPA.

| standard working solution | | main component (IPA) as IS | | | heptadecane as IS | | |
|---|---|---|---|---|---|---|---|
| № | $C_{st}$, mg/ml | $C_{exp(IPA)}$, mg/ml | relative rep., % | relative bias, % | $C_{exp(heptadecane)}$, mg/ml | relative rep., % | relative bias, % |
| 1 | 0,141 | 0,138 | 4,1 | -2,0 | 0,134 | 1,1 | -4,8 |
| 2 | 0,223 | 0,223 | 0,3 | -4,5 | 0,230 | 0,4 | -1,7 |
| 3 | 0,314 | 0,308 | 3,1 | -2,0 | 0,308 | 1,5 | -1,9 |
| 4 | 0,407 | 0,400 | 4,1 | -1,7 | 0,404 | 0,5 | -0,8 |
| 5 | 0,509 | 0,505 | 0,5 | -0,7 | 0,505 | 0,3 | -0,8 |
| 6 | 0,623 | 0,629 | 1,4 | 1,0 | 0,617 | 1,1 | -1,0 |
| 7 | 0,701 | 0,703 | 0,5 | 0,2 | 0,707 | 1,5 | 0,8 |
| 8 | 0,884 | 0,891 | 0,0 | 0,7 | 0.888 | 0,3 | 0,5 |
| 9 | 0,999 | 1,003 | 0,6 | 0,4 | 1,002 | 0,5 | 0,3 |

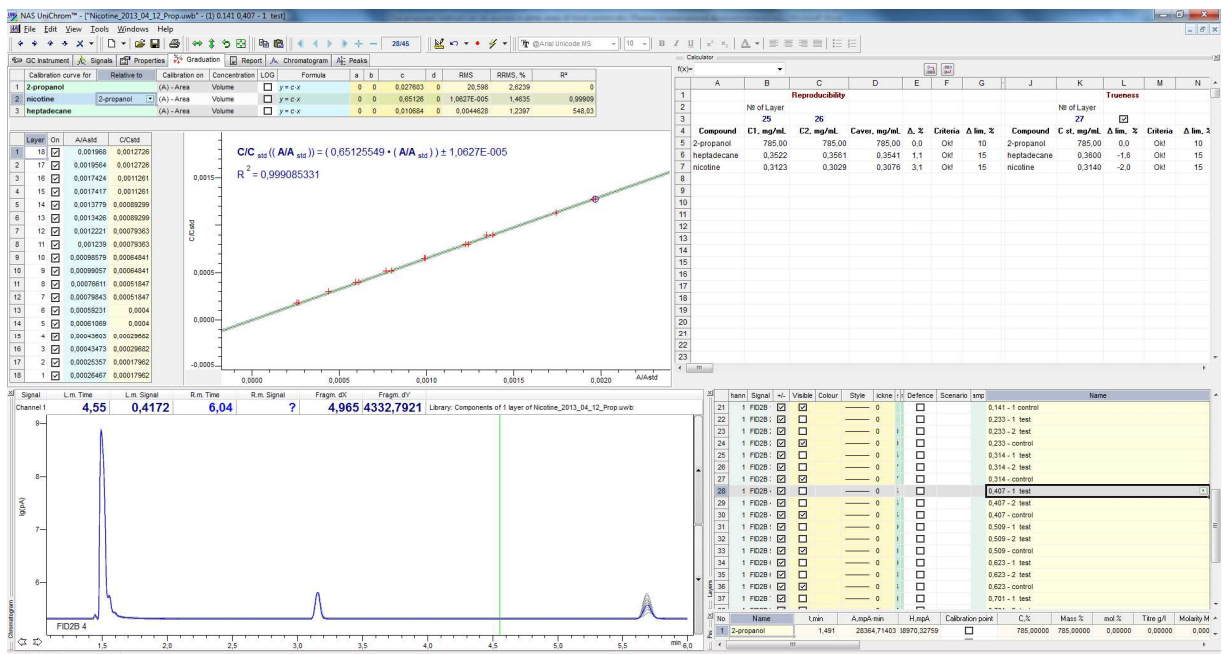

Fig. 11. Experiments with standard working solutions of nicotine in IPA.

In Fig. 11 one can see in the upper left corner calibration characteristics of prepared standard working solutions. In the bottom measured chromatograms with three peaks of IPA, heptadecane and nicotine are presented in logarithmic scale. Peaks of IPA and heptadecane are constant for all nine solutions. Only peak of nicotine is changed because of concentration changes from one solution to another one.

**Conclusions**

The proposed method can be applied in other areas of food control, where it will be the method "Main component as IS". For example, in the determination of nicotine in smoke condensates of cigarettes according to International Standard ISO 10315 "Cigarettes. Determination of nicotine in smoke condensates. Gas Chromatographic method" the main component should be 2-propanol (isopropyl alcohol – IPA). According to the method "Main component as IS" there is no need to add heptadecane as internal standard.

Let us emphasize that analytical, testing and commercial laboratories all over the world may validate proposed new method in their activities, making sure its simplicity, accessibility and effectiveness in everyday practice. This applies to studies of spirit drinks, bioethanol, alcohol production waste, etc. The obtained results show the possibility of developing a new international standard of measurement procedure for determination of volatile compounds in alcohol products. The proposed method gives the following benefits.

1. There is the data accuracy increasing due to exact knowledge of the absolute value of pure ethanol concentration in the sample expressed in milligrams per liter of absolute alcohol.
2. There is no need to add the internal standard substance in the sample.

Independence of the measured results from the strength of the test sample allows us to remove stringent requirements on its minimum volume. As a consequence, the proposed method allows to use certified reference materials CRM LGC5100 Whisky-Congeners as the control samples with a small sample volume, including standard microvials for automatic liquid sampler with internal volume of 2 mL.

*Acknowledgements. We would like to thank Agnieszka Orzeł and Jan Gacko form LGC Standards Sp. z o. o. (Poland) for prompt provision of certified reference material CRM LGC 5100 Whisky-Congeners and New Analytical Systems Ltd. for instrumentation support.*